\documentclass[
    ,final            
  ]
  {aipproc}

\layoutstyle{8x11single}

\usepackage{color}

\begin{document}

\title{Towards an optical potential for rare-earths through coupled~channels}


\author{G. P. A. Nobre}{
  address={National Nuclear Data Center, Brookhaven National Laboratory, Upton, NY 11973-5000, USA}
}

\author{F. S. Dietrich}{
  address={P.O. Box 30423, Walnut Creek, CA, 94598, USA}
}

\author{M. Herman}{
  address={National Nuclear Data Center, Brookhaven National Laboratory, Upton, NY 11973-5000, USA}
}

\author{A. Palumbo}{
  address={National Nuclear Data Center, Brookhaven National Laboratory, Upton, NY 11973-5000, USA}
}

\author{S. Hoblit}{
  address={National Nuclear Data Center, Brookhaven National Laboratory, Upton, NY 11973-5000, USA}
}

\author{D.~Brown}{
  address={National Nuclear Data Center, Brookhaven National Laboratory, Upton, NY 11973-5000, USA}
}


\begin{abstract}
The
coupled-channel theory is a natural way of treating nonelastic channels, in particular those arising from collective excitations, defined by nuclear deformations. Proper treatment of such excitations is often essential to the accurate description of reaction experimental data. Previous works have applied different models to specific nuclei with the purpose of determining angular-integrated cross sections. 
In this work, we present 
an extensive study of the effects of collective couplings and nuclear deformations on integrated cross sections as well as on angular distributions in a consistent manner for neutron-induced reactions on nuclei in the rare-earth region. This specific subset of the nuclide chart was chosen precisely because of a clear static deformation pattern. We analyze the convergence of  the coupled-channel calculations regarding the number of states being explicitly coupled. 
Inspired by the work done by Dietrich \emph{et al.}, a model for deforming the spherical Koning-Delaroche optical potential as function of quadrupole and hexadecupole deformations is also proposed. We demonstrate that the obtained results of calculations for total, elastic and inelastic cross sections, as well as elastic and inelastic angular distributions correspond to a remarkably good agreement with experimental data for scattering energies above around a few MeV.
\end{abstract}

\maketitle


\section{Introduction}
Optical potentials (OP) have been widely used to describe nuclear reaction data by implicitly accounting for the
effects of excitation of internal degrees of freedom and other nonelastic processes. Such optical potentials are
normally obtained through proper parametrization and parameter fitting in order to reproduce specific data sets.
An OP is called global when this fitting process is consistently done for a variety of nuclides.
 
Even though  existing spherically-symmetric OP's  might achieve very good  agreement with experimental data under certain
conditions, as they were specifically designed to do so, they are not reliable at regions without any measurements,
for deformed nuclei, or for the ones away from the stability line. In such circumstances, a more fundamental
approach becomes necessary. 
 
The coupled-channel theory is a natural way of explicitly treating nonelastic channels, in particular those arising
from collective excitations, defined by nuclear deformations. Proper treatment of such excitations is often essential
to the accurate description of reaction experimental data. Previous works have mainly applied different models to specific
nuclei with the purpose of determining angular-integrated cross sections.

There are global spherical OP's that have been fit to nuclei below and above the region of statically deformed
rare-earth nuclei, but these potentials have been viewed as inappropriate for use in coupled-channels calculations,
since they do not account for the loss of flux through the explicitly included inelastic channels.  On the other
hand, a recent paper \cite{Dietrich:2012} shows that scattering from rare earth and actinide nuclei is very near
the adiabatic (frozen nucleus) limit, which suggests that the loss of flux to rotational excitations might be
unimportant.  In this paper we test this idea by performing coupled channel calculations with a global spherical
optical potential by deforming the nuclear radii but making no further adjustments.  We note an alternative
approach (Kuneida \emph{et al.} \cite{Kunieda:2007}), which has attempted to unify scattering from spherical and
deformed nuclei by considering all nuclei as statically deformed, regardless of their actual deformation.

This work corresponds to a preliminary attempt to extend the approach initially presented in Ref.~\cite{NobreND2013}, focusing on angular distributions for the cases of neutron scattered by Gd, Ho, and W nuclei.

\section{Adiabatic model for rare-earths}

Due to the high moment of inertia and consequent low excitation energies of the ground-state band members of the statically deformed nuclei in the rare-earth region, the deformed nuclear configuration may be regarded as ``frozen'' during the scattering. This means that all the internal degrees of freedom not associated with the strong deformation may assumed to be accounted for by a spherical optical potential that describes well the nuclei in the neighboring region, in an adiabatic approach. Therefore, the only channels needed to be treated explicitly (e.\ g., through couple-channel methods) are the ones arising from the static deformation.

The spherical OP that was deformed in our coupled-channel calculations was the global Koning-Delaroche (KD)~\cite{KD}, unmodified except for a small change in the radius parameters to ensure volume conservation when the nucleus is deformed.  Since the KD potential describes scattering from nuclei both above and below the deformed rare earth region very well, we make the assumption that the imaginary potential adequately describes the internal nuclear excitations in the rare earths also.  This is picture is consistent with the adiabatic approximation.  The coupled channel calculations account for the external (rotational) excitations of the target.  These assumptions are tested in the calculations shown in this paper.

The process of deforming a spherical OP to explicitly consider collective excitations within the couple-channel framework is done in the standard way of replacing the radius parameter $R$ in each Woods-Saxon form factor by the angle dependent expression:
\begin{equation}
\label{Eq:DefRadius}
R(\theta)=R_0\left( 1+\sum_\lambda{\beta_\lambda Y_{\lambda0}(\theta)} \right)
\end{equation}
where $R_0$ is the undeformed radius of the nucleus, and $\beta_\lambda$ and $Y_{\lambda0}(\theta)$ are the deformation parameter and spherical harmonic for the multipole $\lambda$, as seen in Ref.~\cite{Krappe1976}, for example. The deformed form factor obtained from Eq.~\ref{Eq:DefRadius} is then expanded in Legendre polynomials numerically.

\begin{figure}
  \includegraphics[height=.3\textheight, clip, trim= 5mm 4mm 5mm 14mm]{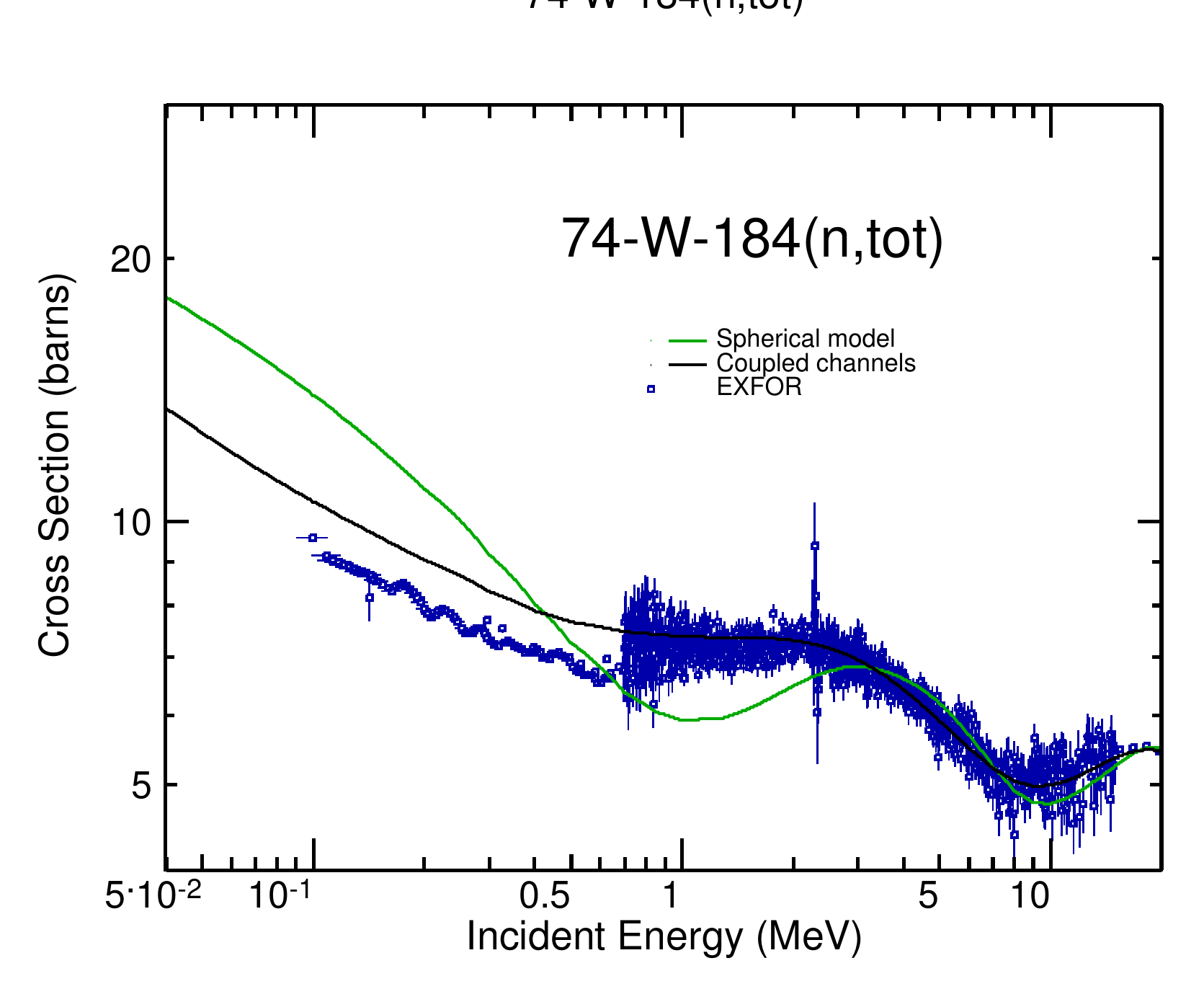}
  \caption{Total cross sections for neutron-induced reaction on $^{184}$W. The black curve corresponds to coupled-channel calculations within our model, while the green curve indicates, for comparison purposes, the result from a spherical model calculation. Experimental data taken from EXFOR \cite{EXFOR}.}
  \label{Fig:Total}
\end{figure}

We use in our calculations the \textsc{Empire} reaction code \cite{Herman:2007,EmpireManual}, in which the direct reaction part is calculated by the code \textsc{Ecis} \cite{Raynal70,Raynal72}. In order to test our model we perform coupled-channel calculations, coupling to the ground state rotational band, for neutron-incident reactions on selected rare-earth nuclei, namely $^{152,154}$Sm, $^{153}$Eu, $^{155,156,157,158,160}$Gd,
$^{159}$Tb, $^{162,163,164}$Dy, $^{165}$Ho,  $^{166,167,168,170}$Er, $^{169}$Tm, $^{171,172,173,174,176}$Yb, $^{175,176}$Lu,
$^{177,178,179,180}$Hf, $^{181}$Ta, and $^{182,183,184,186}$W. All those nuclides have at least 90 neutrons, indicating static deformation, therefore making them suitable candidates for interpolation through the adiabatic limit.     We then compared, as an initial test, the obtained coupled-channel results for total cross sections with plain spherical calculations with the undeformed KD optical potential. In this initial step, only quadrupole deformations were considered, having their values taken from the compilation of of experimental values from Raman \emph{et al.} \cite{Raman}.  The overall result is a dramatic improvement in the agreement  with experimental data, in particular in the lower neutron-incident energies. Fig.~\ref{Fig:Total} clearly illustrates this good improvement in the case of $^{184}$W.

\begin{figure}
  \includegraphics[height=.28\textheight, clip, trim=   5mm 3mm 5mm 14mm]{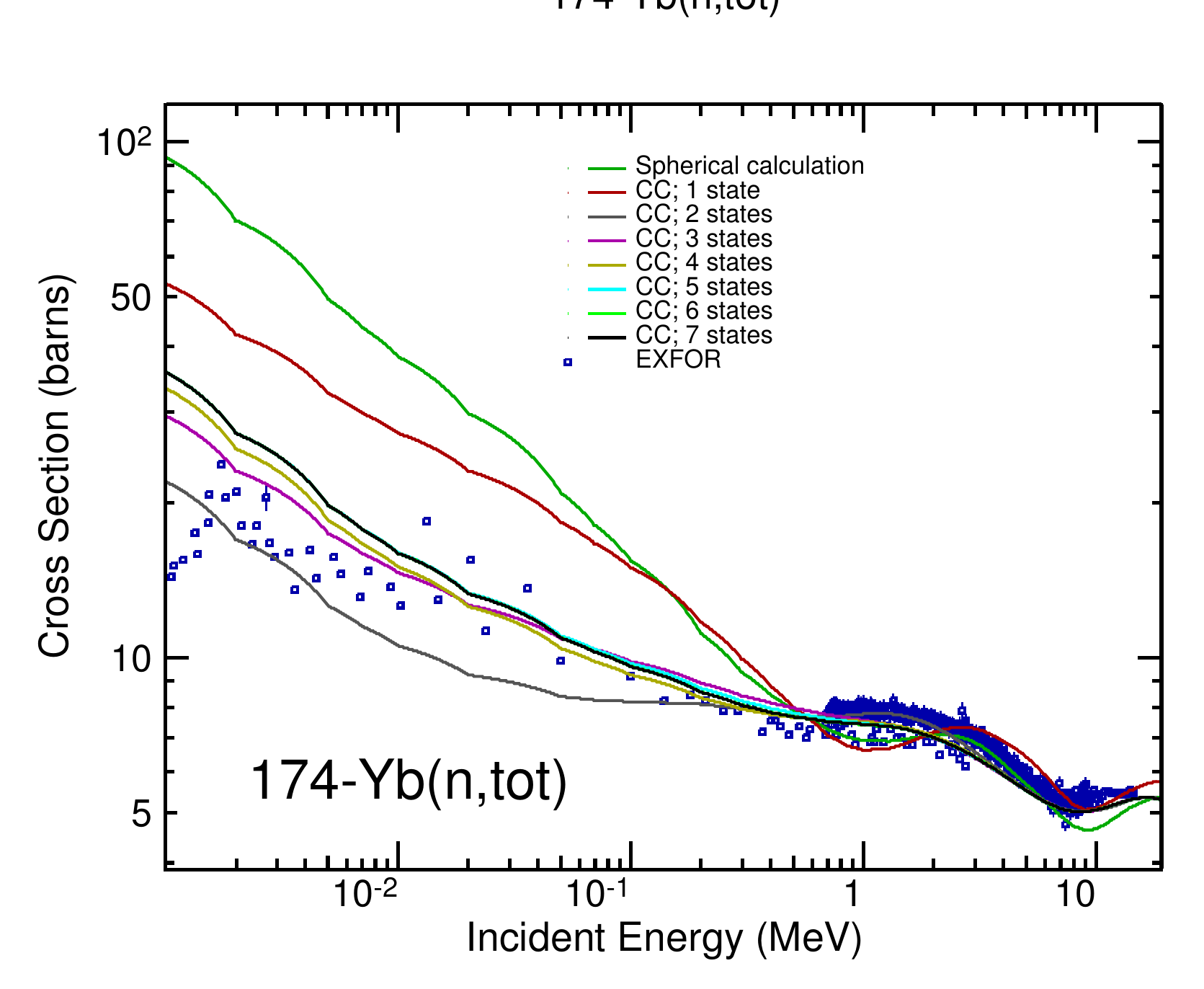} \hspace{1mm}
  \includegraphics[height=.28\textheight, clip, trim= 15mm 3mm 5mm 14mm]{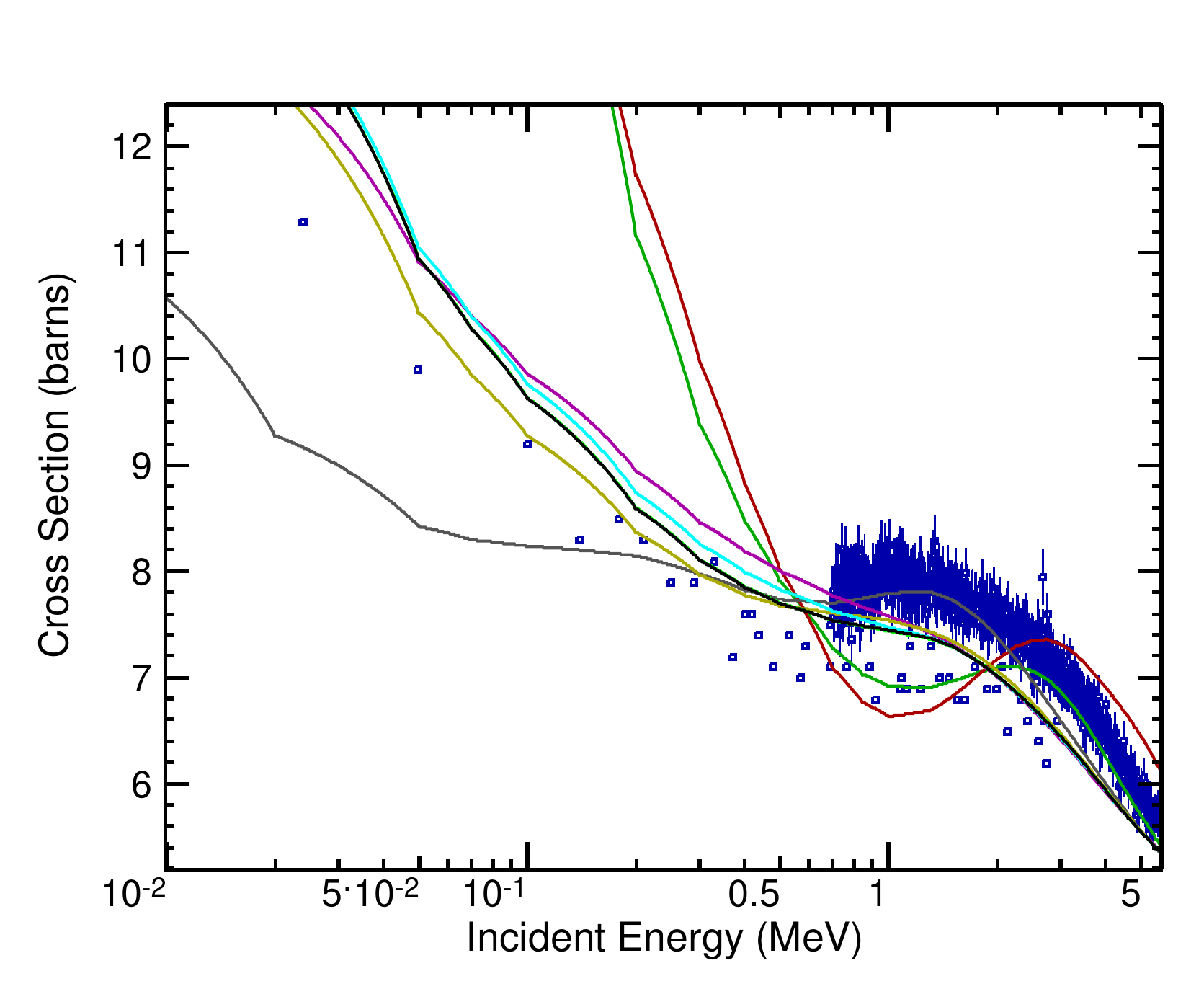}
  \caption{Total cross sections for incident neutrons on $^{174}$Yb.  Right panel corresponds to the same graph as the one on the left panel, but expanded to allow better visualization of region with neutron incident energy between 10 keV and $\sim$ 5 MeV. Dark green curves correspond to calculations within the spherical model, plotted as reference, while other curves are the results obtained by coupling different number of excited states from the ground state rotational band. Experimental data taken from EXFOR \cite{EXFOR}.}
  \label{Fig:Convergence}
\end{figure}

In carrying out the calculations, it is important to couple a sufficient number of rotational states to achieve convergence, and we have carried out tests to ensure this. In Fig.~\ref{Fig:Convergence} we show an example of such analysis for the case of $^{174}$Yb total cross section. On the left panel of Fig.~\ref{Fig:Convergence}  one can see that the couple-channel calculations gradually converge as more excited states are coupled. Even though convergence is overall reached when 5 states are coupled, coupling to a sixth state in the specific region of incident energies between 10 keV and $\sim$ 5 MeV seems necessary. The right panel of Fig.~\ref{Fig:Convergence} shows the same graph as the left panel but with an expanded scale in this region so the effect of this additional coupling can be better visualized.

\subsection{Compound-nucleus observables}

\begin{figure}
  \includegraphics[height=.25\textheight, clip, trim=     4mm 3mm 5mm 3mm]{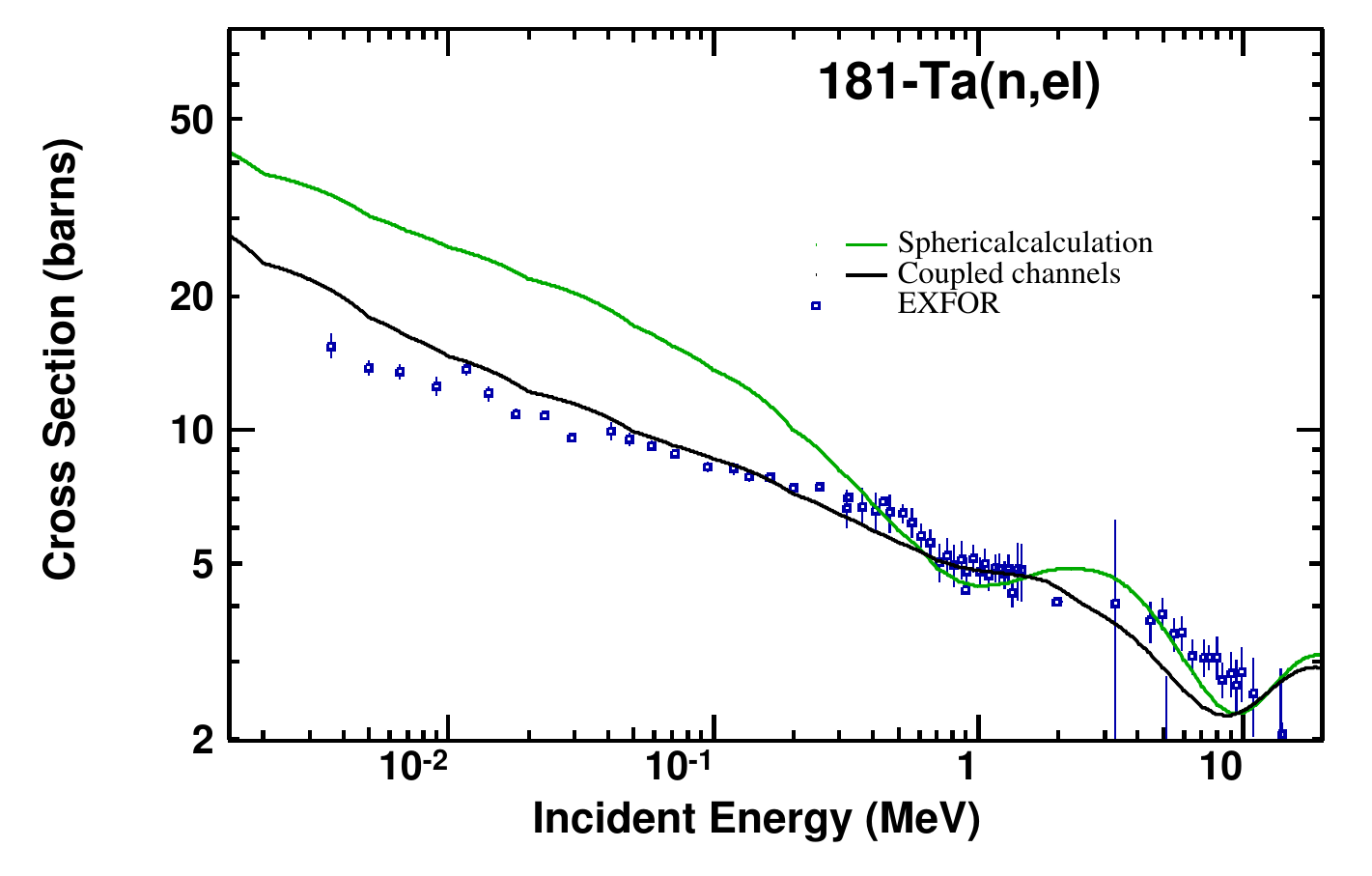} \hspace{-1.8mm}
  \includegraphics[height=.25\textheight, clip, trim=   24mm 3mm 5mm 3mm]{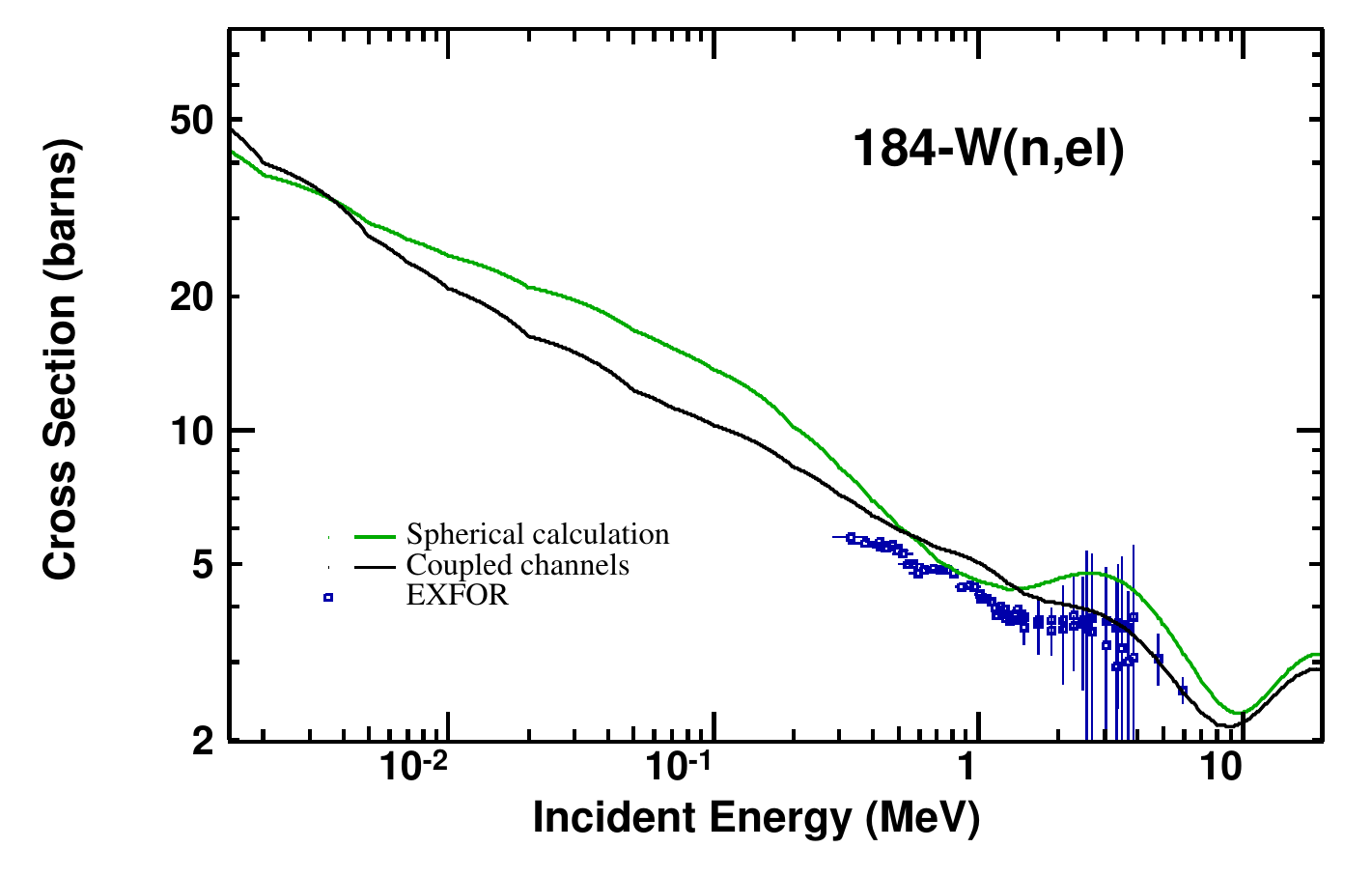}
  \caption{Angle-integrated elastic cross sections for the case of  $^{181}$Ta (left panel) and $^{184}$W (right panel) targets. Black curves correspond to predictions by our coupled-channel model while green curves were obtained by spherical model calculations. Experimental data taken from EXFOR \cite{EXFOR}.}
  \label{Fig:Elastic}
\end{figure}

After the initial success in describing direct-reaction quantities, such as total cross sections, we analyzed the model predictions for observables that depend also on the compound-nucleus decay. The models adopted to describe the emissions from the compound nucleus were basically default options in \textsc{Empire} code, which means standard Hauser-Feshbach model with properly parametrized Enhanced Generalized Superfuid Model (EGSM) level densities \cite{fade}, modified Lorentzian distribution (version 1) for $\gamma$-ray strength functions \cite{plu01,plu02,plu03}, width fluctuation correction implemented up to 3 MeV in terms of the HRTW approach \cite{HRTW,HHM}, and with transmission coefficients for the inelastic outgoing channels also calculated within coupled-channel approach (spherical KD also used in outgoing channels). Pre-equilibrium was calculated within the exciton model~\cite{Griffin:66}, as based on the solution of the
master equation~\cite{Cline:71} in the form proposed by Cline~\cite{Cline:72}
and Ribansky~\cite{Ribansky:73} (using \textsc{Pcross} code \cite{Herman:2007,EmpireManual}) with mean free path multiplier set to 1.5.
 In Fig.~\ref{Fig:Elastic}, as an example, we compare with experimental data the angle-integrated elastic cross sections for incident neutrons on $^{181}$Ta and $^{184}$W obtained by our coupled-channel calculations. As it may be seen in Fig.~\ref{Fig:Elastic}, while the spherical-model calculations fail to reproduce the measured cross sections, our coupled-channel results are in a very good agreement with experimental data.

\begin{figure}
  \includegraphics[height=.22\textheight, clip, trim=     27mm 28mm 32mm 119mm]{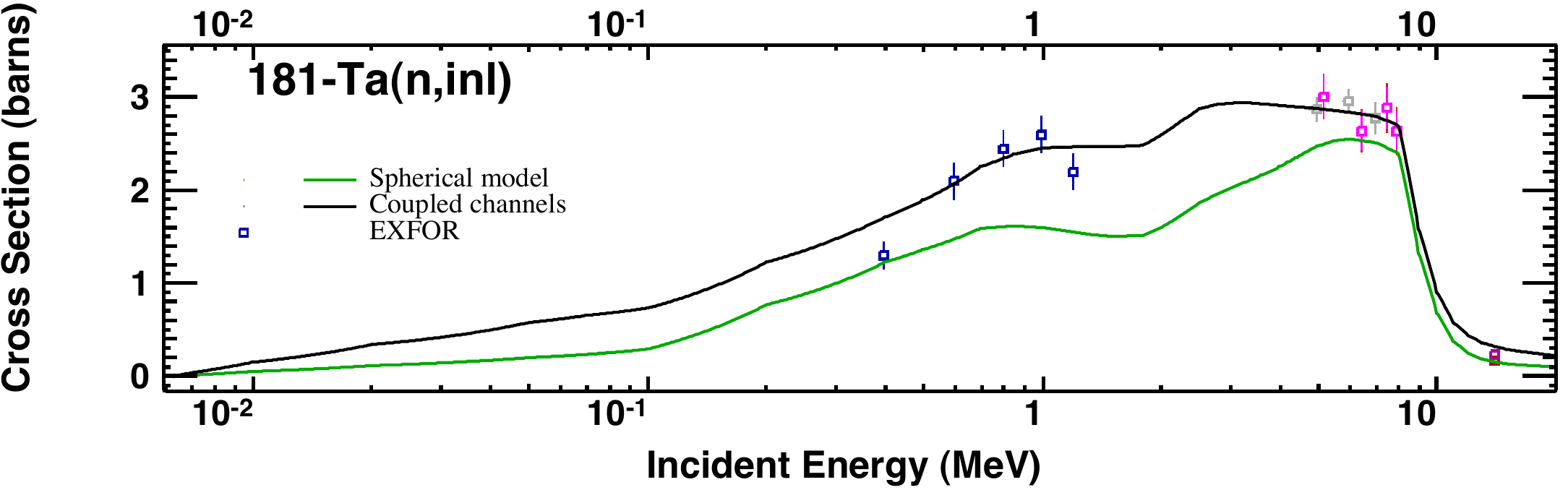}
  \caption{Angle-integrated inelastic cross sections for $^{181}$Ta. Black curves correspond to predictions by our coupled-channel model while green curves were obtained by spherical model calculations. Experimental data taken from EXFOR \cite{EXFOR}.}
  \label{Fig:Inel}
\end{figure}

Fig. \ref{Fig:Inel} shows the total inelastic cross section in the case of neutrons scattered by a $^{181}$Ta target. Again, our coupled-channel model describes well the observed experimental data.

\section{Angular Distributions}

A more careful analysis of differential cross-section experimental data proved necessary due to the large amount of angular distribution data available in the literature, and also because some measurements do not contain pure elastic isotopic data. It is quite common for experiments measuring elastic angular distributions for  rare-earth nuclei to be unable to separate inelastic contributions due to the low-lying excitation energies of their rotational states. In such cases, measured data correspond actually to ``quasi-elastic'' angular distributions, and the  calculated elastic and inelastic differential cross sections have to added up together accordingly for appropriate comparison. In addition, some experiments were done using the natural form of the element, rather than the isotope-specific one.

For these reasons, application of the coupled-channel model for angular distributions was focused on three elements only: Gadolinium, Holmium, and Tungsten. Those three elements were  chosen because the lighter and heavier ones are close to the border of the rare-earth region, while the other is roughly in the middle.

\subsection{Volume Conservation}
 
When an originally spherical configuration assumes a deformed shape, defined by quadrupole and hexadecupole
deformation parameters $\beta_2$ and $\beta_4$, respectively, the volume and densities are not conserved. In Ref.~\cite{Bang:1980}, a method
to ensure volume conservation was proposed, corresponding to applying  a correction to the reduced radius $R_0$, of the form:
\begin{equation}
 R'_{0}=R_0\left(  1-\sum^{}_{\lambda}{\beta_{\lambda}^{2}/4\pi}\right) ,
\label{Eq:radius}
\end{equation}
where terms of the order of $\beta_{\lambda}^{3}$ and higher have been discarded. Ref.~ \cite{NobreND2013} tested the effects of such correction, showing that it is not negligible and seems to bring the integral and differential cross-section calculations to a slightly better agreement with
the experimental data. Therefore, in the following calculations of angular distributions, we decided to implement the radial corrections calculated from Eq.~\ref{Eq:radius}, as it should correspond to a more realistic modeling of the deformed nuclei.

\subsection{Initial results}

In this work we present preliminary results of angular distributions for $^{158}$Gd and $^{184}$W. In Ref.~\cite{Herman-CNR} one can also find preliminary results for quasi-elastic differential cross sections for the $^{165}$Ho target within the same coupled-channel model.

\begin{figure}
\begin{minipage}[c][11cm]{\textwidth}
\vspace*{\fill}
\centering
  \includegraphics[height=.5\textheight]{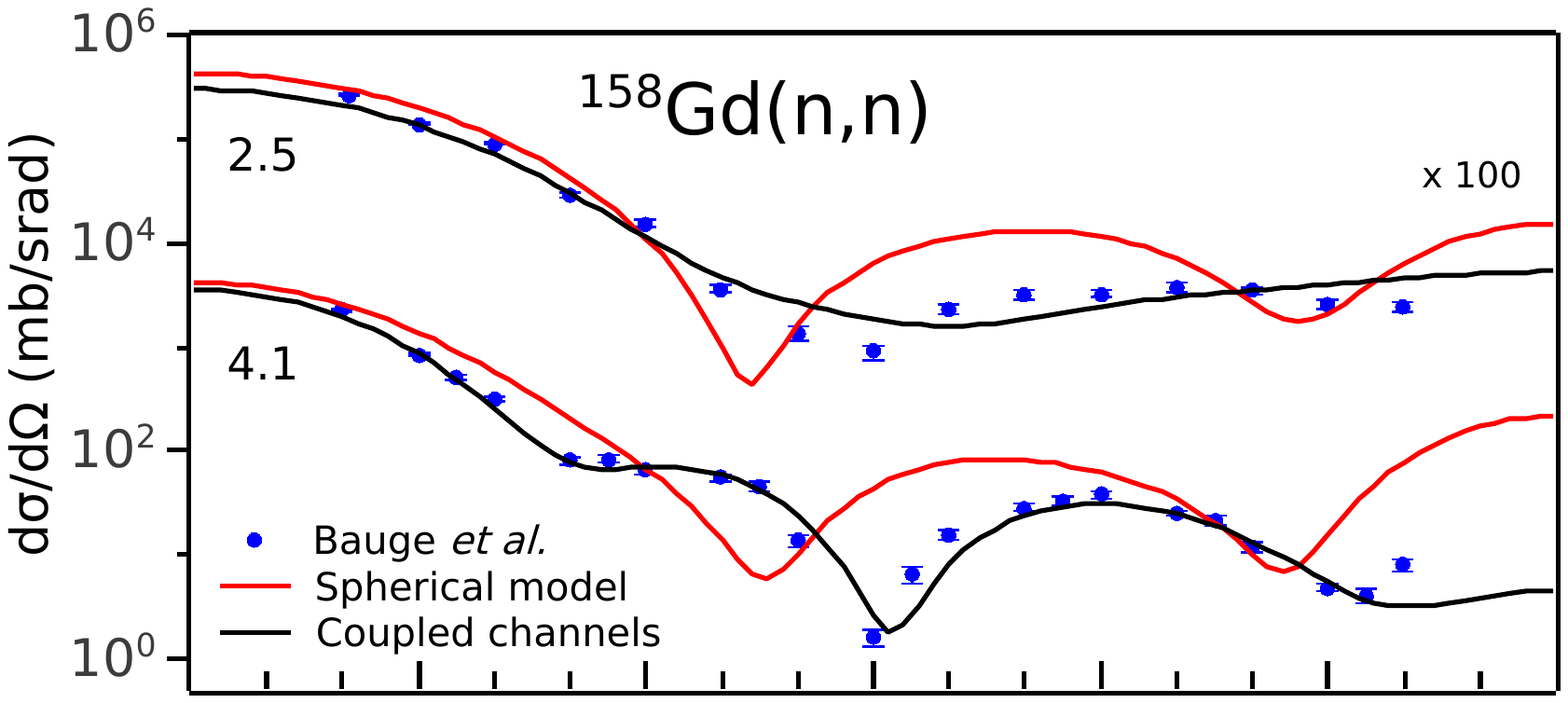} \\ \vspace{-7.40cm}
  \includegraphics[height=.5\textheight]{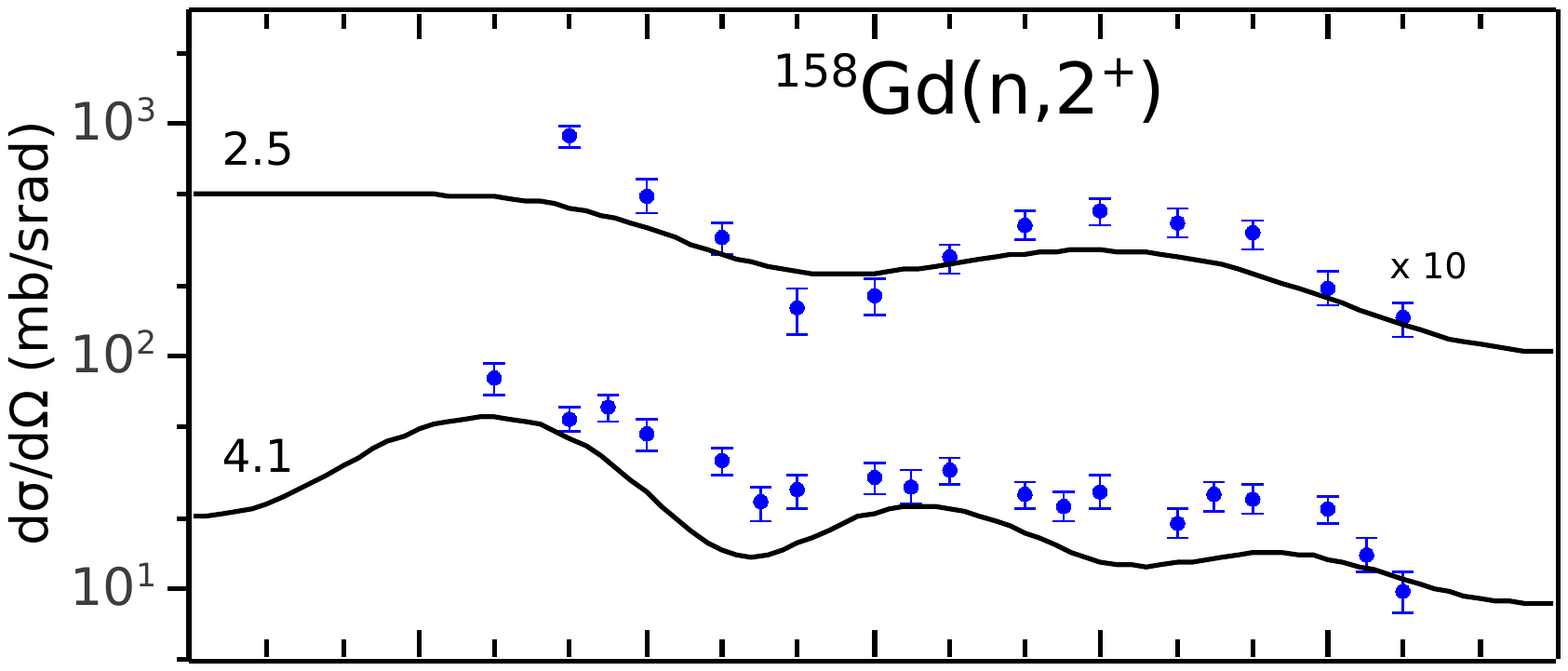} \\  \vspace{-7.40cm} 
  \includegraphics[height=.5\textheight]{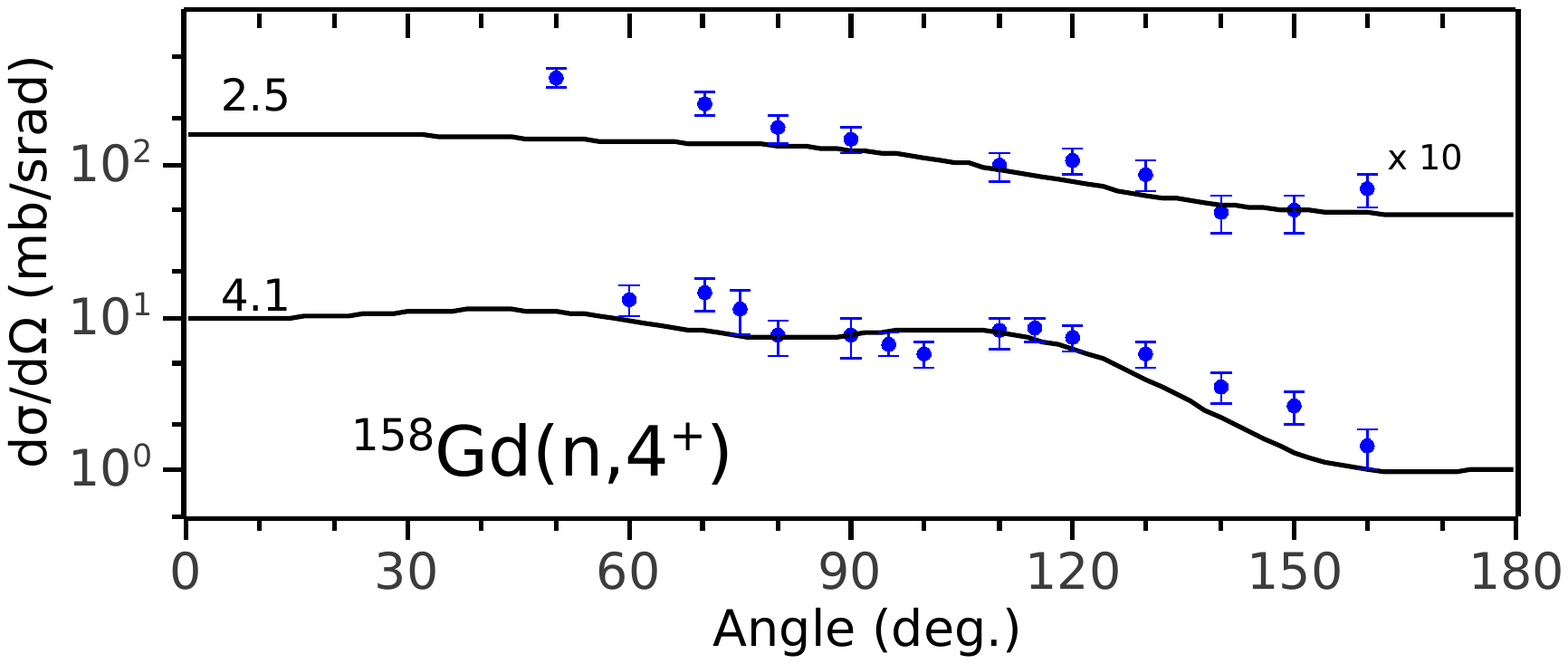} \\  \vspace{1cm} 
  \caption{Angular distributions for elastic (upper panel) and first and second inelastic  (middle and bottom panels, respectively)  channels for the neutron-induced reaction on $^{158}$Gd. Black curves correspond to predictions by our coupled-channel model while red curves were obtained by spherical model calculations. Numbers on the left indicate, in MeV, the values of incident energy at which the cross sections were measured. Experimental data taken from Bauge \emph{et al.} \cite{Bauge:2000}.}
  \label{Fig:Gd158}
\end{minipage}  
\end{figure}

Fig.~\ref{Fig:Gd158}  presents the predictions of our model for the elastic angular distribution for $^{158}$Gd. In the same figure it is also shown the differential cross sections for the first 2$^+$ and 4$^+$ states, which have excitation energy of 79.5~keV and 261.5~keV, respectively. The values used for the deformation parameters were $\beta_{2}=0.348$ \cite{Raman} and $\beta_{4}=0.056$ \cite{Smith:2004}. From Fig.~\ref{Fig:Gd158}, it can be clearly seen that our model succeeds in reproducing very well the observed elastic differential cross section (upper panel), especially when compared with the results from the spherical model calculation. The couple-channel agreement is still not perfect but, considering the fact that no adjustment of parameters was performed, this corresponds to an impressive preliminary result.

Regarding the predictions of our coupled-channel model for the angular distributions of the first two excited states, as shown in Fig.~\ref{Fig:Gd158} (middle and bottom panels), even though the agreement with experimental data is not as good as in the elastic scattering case, they still describe reasonably well the measured data, in particular their shape.

\begin{figure}
\centering
 \begin{minipage}[c]{.5\textwidth}
\vspace*{\fill}
  \includegraphics[height=.430\textheight, clip, trim= 5mm -0mm -2mm -2mm]{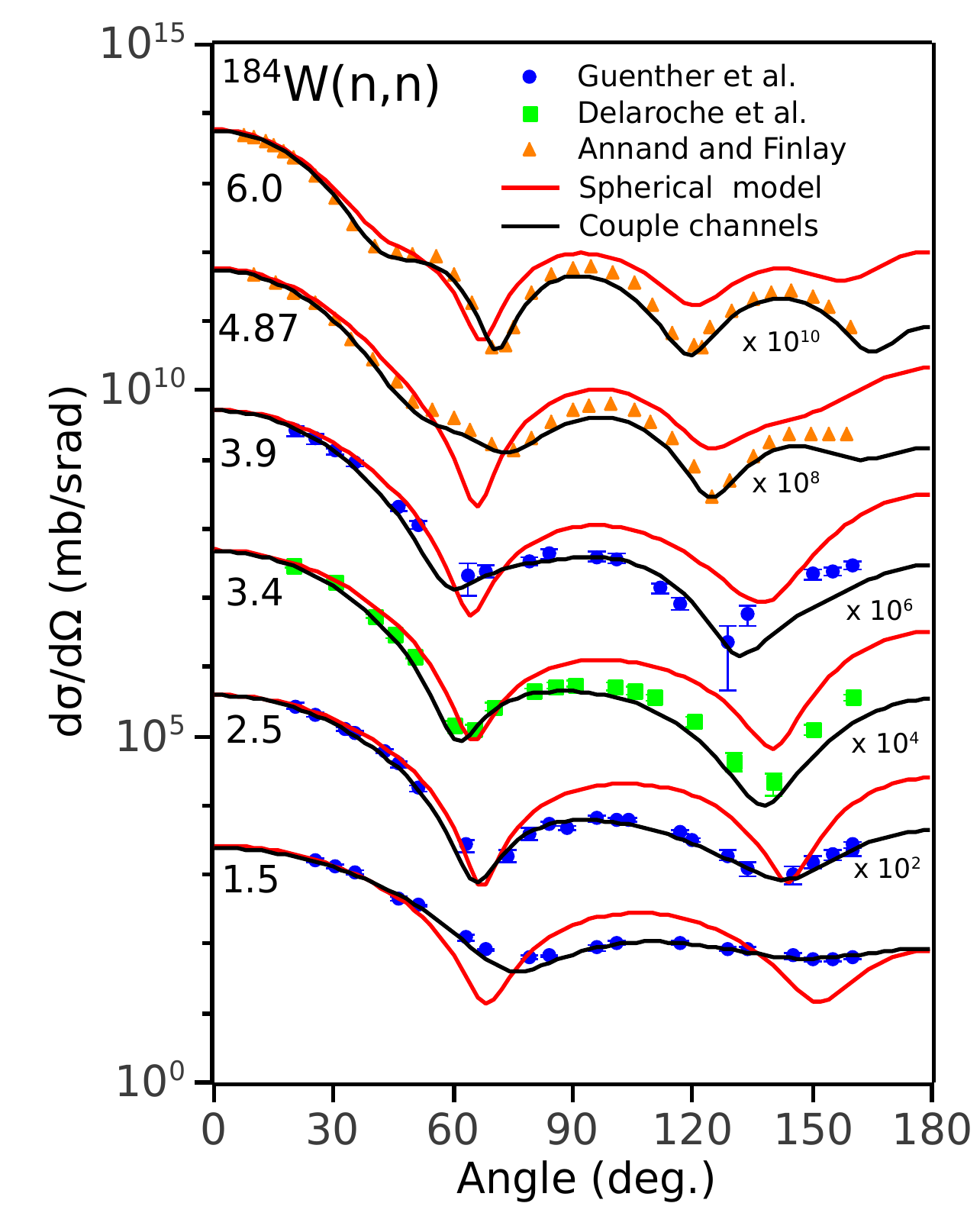}
  \end{minipage}  \hspace{-8mm}
   \begin{minipage}[c]{.5\textwidth}
\vspace*{\fill}
 \hspace{-0.5mm}  \includegraphics[scale=.349, clip, trim= 9mm 1mm 0mm 0mm]{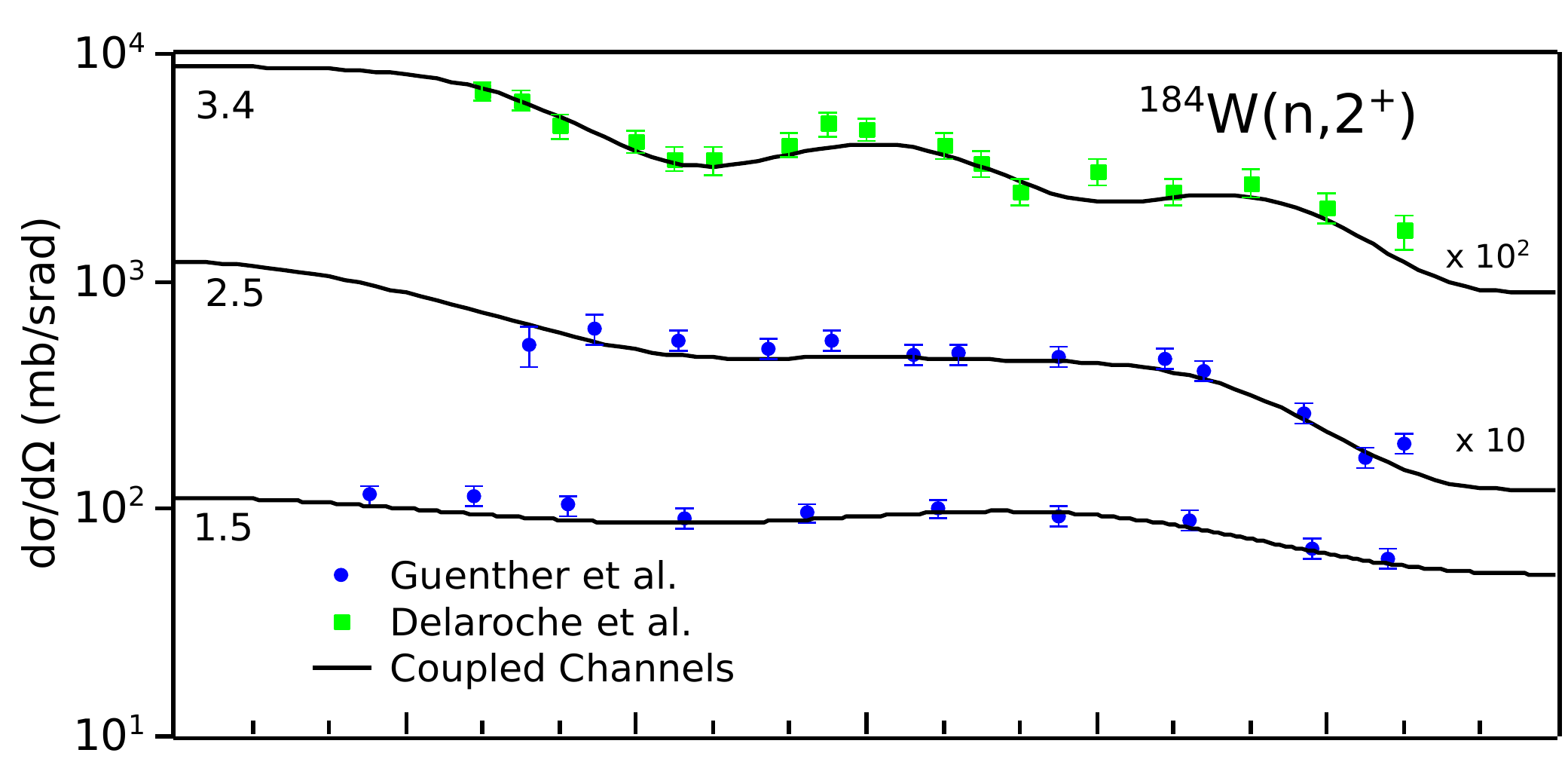}  \\ \vspace{-0cm}
  \includegraphics[scale=.360, clip, trim= 13mm 1mm 0mm 16mm]{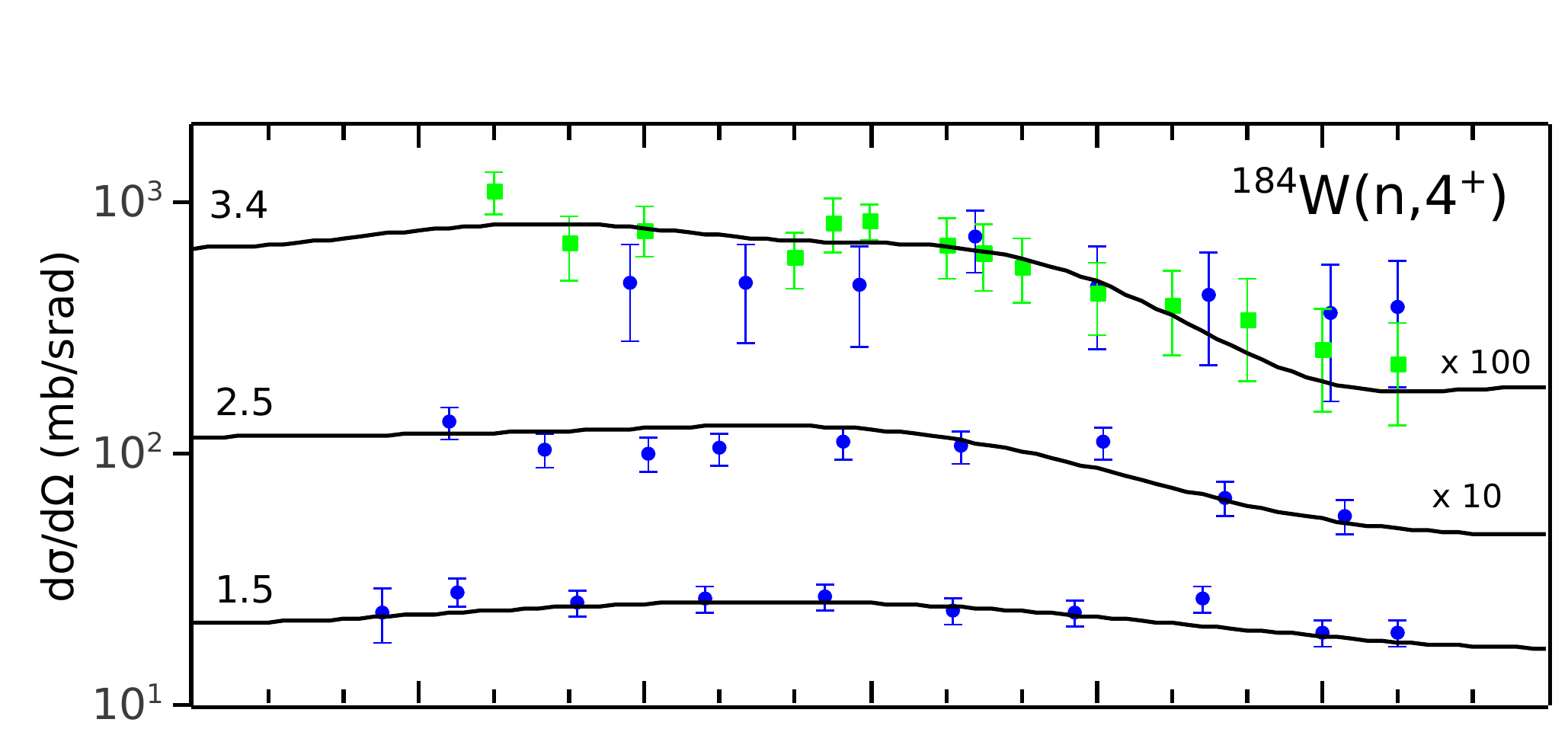}  \\ \vspace{0.cm}
  \includegraphics[scale=.363, clip, trim= 14.5mm 1mm 0mm 9mm]{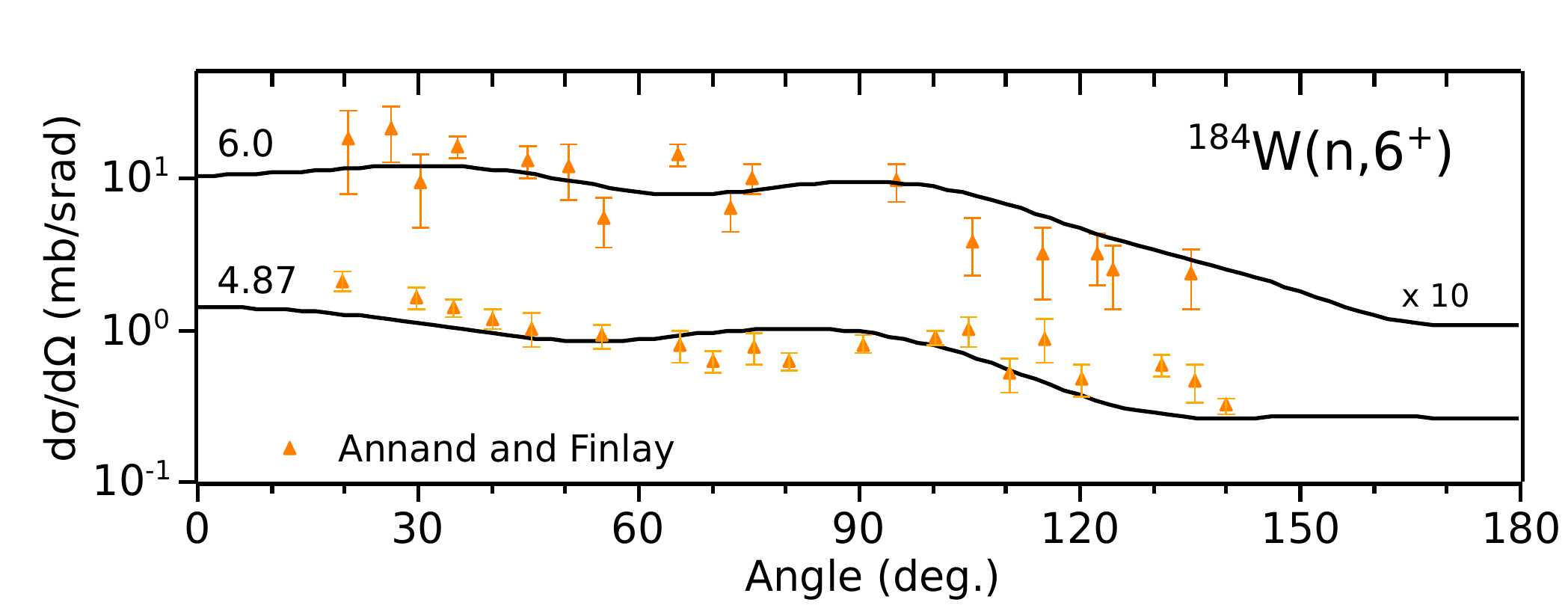}  
    \end{minipage} 
  \caption{Angular distributions for elastic (left panel) and first, second and third inelastic  (right panels) channels for the neutron-induced reaction on $^{184}$W. Black curves correspond to predictions by our coupled-channel model while red curves were obtained by spherical model calculations. Numbers on the left of each graph indicate, in MeV, the values of incident energy at which the cross sections were measured. Experimental data taken from Refs.~\cite{Guenther:1982,Delaroche:1981,Annand:1985}.}
\label{Fig:W184}
\end{figure}

In Fig. \ref{Fig:W184} we present the angular-distribution results obtained through our coupled-channel model for neutrons scattered by a $^{184}$W target nucleus for the elastic (left panel) and first three inelastic (right panels) channels . The excitation energies of the 2$^+$, 4$^+$, and 6$^+$ states for which differential cross sections are shown are, respectively, 111.2~keV, 364.1~keV, and 748.3~keV. We adopted the values $\beta_{2}=0.236$ \cite{Raman} and $\beta_{4}=-0.080$ \cite{Hendrie:1968} for the quadrupole and hexadecupole deformation parameters, respectively.

Similarly as for the case of $^{158}$Gd, we observe that we achieve a very good description of elastic and inelastic differential data for $^{184}$W, especially if we compare with the spherical-model calculations. Particularly noteworthy is the good description of the angular distribution for the 6$^+$, even though we do not couple directly from ground state to that particular state, that means we have assumed for now $\beta_{6}=0$.

\section{Conclusion}

In this work we have demonstrated that deforming the spherical Koning-Delaroche optical potential 
and using it in coupled channels calculations without further modification provides very good
results in the description of neutron-induced reactions on the rare-earths, despite the fact this potential was not
designed to describe such deformed nuclei. We apply a correction in the reduced radius of the target-nuclei to ensure
volume conservation when deforming an originally spherical configuration. 
We achieved a good description of experimental data not only for optical-model observables (such as total cross sections,
elastic and inelastic angular distributions), but also for those obtained through compound-nucleus formation (such as
total elastic and inelastic cross sections). These good results are consistent with the insight gained from
Ref. \cite{Dietrich:2012} that the scattering is very close to the adiabatic limit. Similarly impressive results were obtained when we preliminarily attempted to describe  observed elastic and inelastic angular distributions.  Although the presented results are
not perfect, this simple method corresponds to a good, consistent and general first step towards an optical potential
capable of fully describing the rare-earth region, filling the current lack of optical model potentials in this important region.


\begin{theacknowledgments}
The work at Brookhaven National Laboratory was sponsored by the Office of Nuclear
Physics, Office of Science of the U.S. Department of
Energy under Contract No. DE-AC02-98CH10886 with
Brookhaven Science Associates, LLC.
\end{theacknowledgments}



\bibliographystyle{aipproc}   

\bibliography{NobreGPA}

\IfFileExists{\jobname.bbl}{}
 {\typeout{}
  \typeout{******************************************}
  \typeout{** Please run "bibtex \jobname" to optain}
  \typeout{** the bibliography and then re-run LaTeX}
  \typeout{** twice to fix the references!}
  \typeout{******************************************}
  \typeout{}
 }

\end{document}